\documentclass[aps,preprint]{revtex4}%
\usepackage{amsfonts}
\usepackage{amsmath}
\usepackage{amssymb}
\usepackage{graphicx}%
\begin{document}
\title{Thermodynamic extremality relations in the massive gravity}
\author{Deyou Chen}
\email{deyouchen@hotmail.com}
\affiliation{School of Science, Xihua University, Chengdu 610039, China}

\author{Jun Tao}
\email{taojun@scu.edu.cn}
\author{Peng Wang}
\email{pengw@scu.edu.cn}
\affiliation{Center for Theoretical Physics, College of Physics, Sichuan University, Chengdu 610064, China}

\begin{abstract}
{A universal relation between the leading correction to the entropy and extremality was gotten in the work of Goon and Penco. In this paper, we extend this work to the massive gravity and investigate thermodynamic extremality relations in a topologically higher-dimensional black hole. A rescaled cosmological constant is added to the action of the massive gravity as a perturbative correction. This correction modifies the extremality bound of the black hole and leads to the shifts of the mass, entropy, etc. The Goon-Penco relation is gotten. Regarding the cosmological constant as a variable related to pressure, we get the thermodynamic extremality relations between the mass and pressure, charge, parameters $c_i$ by accurate calculations, respectively. Finally, these relations are verified by a triple product identity, which shows that the universal relation exists in black holes.}
\end{abstract}
\maketitle
\tableofcontents

\section{Introduction}

The string landscapes formed by effective quantum field theories are broad and complex. However, there are some theories that look self-consistent but not compatible with the string theory. Thus, the swampland program was put forward \cite{CV,AMNV,EP,BCV}. Its aim is to find the subset of the infinite space in effective field theories arisen at low energies from quantum gravity theories by specific constraints. These constraints were first proposed in \cite{CV}. As one of the constraints, the weak gravity conjecture (WGC) has attracted people's attention. It asserts that for the lightest charged particle along the direction of some basis vectors in charge space, the charge-to-mass ratio is larger than for extremal black holes \cite{AMNV}. This conjecture shows that the extremal black holes are allowed to decay.

A proof to the WGC is that it is mathematically equivalent to a certain property of a black hole entropy. In \cite{CLR1}, the authors introduced the higher-derivative operators to the action to compute the shift of the entropy. Due to these operators, the extremality condition of the black hole is modified and the mass and entropy are shifted. They derived the relation between the ratio of charge-to-mass and the entropy shift, $q/m -1 \propto \Delta S$, where $\Delta S >0$. The charge-to-mass ratio asymptotes to unity with the increase of the mass. Thus, the large extremal black hole is unstable and decays to a smaller extremal black hole with the charge-to-mass ratios greater than unity. This phenomenon satisfies the requirement of the WGC. Subsequently, the WGC behavior was found in the four-dimensional rotating dyonic black hole and other spacetimes \cite{CLR2,JM}. Other researches on the WGC are referred to \cite{CHNW,ACS,LLW1,LLW2,KMP,KMP1,KMP2,LNS,MRV,CCLOR,COR,RS,HRR,BH,GI,HNS} and references therein.

In the recent work \cite{GP}, Goon and Penco derived a universal extremality relation by the perturbative corrections to the free energy of generic thermodynamic systems. This relation takes the form

\begin{eqnarray}
\frac{\partial M_{ext}(\overrightarrow{\mathcal{Q}},\epsilon)}{\partial \epsilon} = \lim_{M \to M_{ext}(\overrightarrow{\mathcal{Q}},\epsilon)} - T\left(\frac{\partial S(M,\overrightarrow{\mathcal{Q}},\epsilon)}{\partial \epsilon}\right)_{M,\overrightarrow{\mathcal{Q}}},
\label{eq1.1}
\end{eqnarray}

\noindent where $M_{ext}(\overrightarrow{\mathcal{Q}},\epsilon)$ and $S(M,\overrightarrow{\mathcal{Q}},\epsilon)$ are the extremality mass and entropy, respectively. Both of them are $\epsilon-$dependent and $\epsilon$ is a control parameter in front of the free energy. $\overrightarrow{\mathcal{Q}}$ are additional quantities in the thermodynamic systems other than the mass. The above relation can be interpreted as a comparison between states in the classical and corrected theories. Meanwhile, an approximation relation  $\Delta M_{ext}(\overrightarrow{\mathcal{Q}}) \approx -T_0(M,\overrightarrow{\mathcal{Q}}) \Delta S(M,\overrightarrow{\mathcal{Q}})|_{M\approx M_{ext}^0(\overrightarrow{\mathcal{Q}})}$ was gotten, where $\Delta M_{ext}(\overrightarrow{\mathcal{Q}})$ and $\Delta S(M,\overrightarrow{\mathcal{Q}})$ are the leading order corrections to the extremal bound and to the entropy of a state at the fixed mass and $\overrightarrow{\mathcal{Q}}$, respectively. $M_{ext}^0$ is the mass in the extremal case without corrections. The result shows that the mass of the perturbed extremal black hole is less than that of the unperturbed one with the same quantum numbers, if $\Delta S>0$, which implies that the perturbation decreases the mass of the extremal black hole. Therefore, the WGC-like behavior exists in the extremal black hole. In particular, the Goon-Penco relation (\ref{eq1.1}) was verified in the AdS-Reissner-Nordstr$\ddot{o}$m black hole by rescaling the cosmological constant as a perturbative correction. The approximation relation was also checked by the higher-derivative operators introduced in the action.

To further explore the WGC behavior and the Goon-Penco relation, people studied the thermodynamic corrections in the specific spacetimes by introducing the higher-derivative operators or perturbative parameters \cite{CJLM,WYL}. The Goon-Penco relation was confirmed and other extremality relations were gotten. In \cite{CJLM}, Cremonini et al. computed the four-derivative corrections to thermodynamic quantities in the higher-dimensional AdS-Reissner-Nordstr$\ddot{o}$m black hole and found the extremality relation between the mass and charge,

\begin{eqnarray}
\lim_{T \to 0}\left(\frac{\partial M}{\partial \epsilon} \right)_{Q,T}= \lim_{T \to 0} - \Phi \left(\frac{\partial Q }{\partial \epsilon}\right)_{M,T}.
\label{eq1.2}
\end{eqnarray}

\noindent Extended this work to rotating anti-de Sitter spacetimes, Liu et al. derived the extremality relation between the mass and angular momentum in the BTZ and Kerr anti-de Sitter spacetimes \cite{WYL},

\begin{eqnarray}
\left(\frac{\partial M_{ext}}{\partial \epsilon}\right)_{J,l}  = \lim_{M \to M_{ext}} - \Omega\left(\frac{\partial J}{\partial \epsilon}\right)_{M,S,l}.
\label{eq1.3}
\end{eqnarray}

\noindent The relations (\ref{eq1.2}) and (\ref{eq1.3}) are the extensions of the Goon-Penco relation (\ref{eq1.1}). These relations will shed light in theories of quantum gravity.

In this paper, we extend the work of \cite{GP} to the massive gravity, and investigate the extremality relations between the mass and pressure, entropy, charge, parameters $c_i$ of a charged topological black hole in the higher-dimensional spacetime, respectively. Einstein's general relativity (GR) is a low energy effective theory. The UV completeness requires that GR be modified to meet physical descriptions in the high energy region. The massive gravity is a straightforward modification to GR. We introduce a perturbative correction by adding a rescaled cosmological constant to the action of the massive gravity. This scenario is different from that in \cite{GP} where the cosmological constant was directly rescaled in the action and consistent with that in \cite{WYL}. In our investigation, the cosmological constant is regarded as a variable related to pressure \cite{KRT,KRT1,LPVP,KM1}. Its conjugate quantity is a thermodynamic volume. The black hole mass is naturally interpreted as an enthalpy. The first reason for this is that the cosmological constant, as a variable, can reconcile the inconsistency between the first law of thermodynamics of black holes and the Smarr relation derived from the scaling method. The second reason is that physical constants, such as the gauge coupling constants, Newtonian constant or cosmological constant arisen as vacuum expectation values are not fixed and vary in the more fundamental theories \cite{CM}.

The rest of this paper is organized as follows. In the next section, the solution of the higher-dimensional black hole in the massive gravity is given and its thermodynamic properties are discussed. In section 3, we introduce a perturbative correction into the action and derive the extremality relations between the mass and pressure, entropy, charge, parameters $c_i$, respectively. Section 4 is devoted to our discussion and conclusion.

\section{The black hole solution in the massive gravity}

The action for an $(n + 2)$-dimensional massive gravity is \cite{DV}

\begin{eqnarray}
\mathcal{S} &=& \frac{1}{16\pi}\int{dx^{n+2}\sqrt{-g}\left[R +\frac{n(n+1)}{l^2}-\frac{F^2}{4}+m^2\sum_{i=1}^4{c_iu_i(g,f)}\right]},
\label{eq2.1}
\end{eqnarray}

\noindent where the terms including $m^2$ are the massive potential associate with graviton mass, $f$ is a fixed symmetric tensor called as the reference metric, $c_i$ are constants, and $u_i$ are symmetric polynomials of the eigenvalues of the $(n+2)\times (n+2)$ matrix $\mathcal{K}_{\nu}^{\mu}=\sqrt{f^{\mu \alpha}g_{\alpha \nu}}$:

\begin{eqnarray}
u_1 &=& [\mathcal{K}], \quad u_2=[\mathcal{K}]^2-[\mathcal{K}^2], \quad u_3=[\mathcal{K}]^3-3[\mathcal{K}][\mathcal{K}^2]+2[\mathcal{K}^3],\nonumber\\
u_4 &=& [\mathcal{K}]^4-6[\mathcal{K}^2][\mathcal{K}]^2+8[\mathcal{K}^3][\mathcal{K}]+3[\mathcal{K}^2]^2-6[\mathcal{K}^4].
\label{eq2.2}
\end{eqnarray}

\noindent The square root in $\mathcal{K}$ denotes $(\sqrt{A})_{\nu}^{\mu}(\sqrt{A})_{\lambda}^{\nu}=A_{\lambda}^{\mu}$ and $[\mathcal{K}]= \mathcal{K}_{\mu}^{\mu}$.

The solution of the charged black hole with the spacetime metric and reference metric is given by \cite{CHPZ}

\begin{eqnarray}
ds^2 = -f(r)dt^2 + \frac{1}{f(r)}dr^2 + r^2 h_{ij}dx^idx^j,
\label{eq2.3}
\end{eqnarray}

\begin{eqnarray}
f_{\mu\nu}=diag(0,0,c_0^2h_{ij}),
\label{eq2.4}
\end{eqnarray}

\noindent where

\begin{eqnarray}
f(r)&=& k+\frac{r^2}{l^2}-\frac{16\pi M}{n\Omega_nr^{n-1}}+\frac{(16\pi Q)^2}{2n(n-1)\Omega_n^2r^{2(n-1)}}+\frac{c_0c_1m^2r}{n}+c_0^2c_2m^2\nonumber\\
&&+\frac{(n-1)c_0^3c_3m^2}{r} +\frac{(n-1)(n-2)c_0^4c_4m^2}{r^2},
\label{eq2.5}
\end{eqnarray}

\noindent $l^2$ is related to the cosmological constant $\Lambda$ as $l^2 = -\frac{n(n+1)}{2\Lambda }$. $M$ and $Q$ are the mass and charge of the black hole, respectively. $\Omega_n$ is the volume spanned by coordinates $x^i$ and $c_0$ is a positive integral constant. $h_{ij}dx^idx^j$ is the line element for an Einstein space with the constant curvature $n(n-1)k$. $k = 1$, $0$ or $-1$ denote a spherical, Ricci flat or hyperbolic topology black hole horizons, respectively. The thermodynamics in the extended phase space of the massive gravity were studied in \cite{DYC,XCH1,XCH2,XCH3}. The event horizon $r_+$ is determined by $f (r)=0$. The Hawking temperature is

\begin{eqnarray}
T &=& \frac{1}{4\pi r_+} \left[\frac{(n+1)r_+^2}{l^2}+\frac{(16\pi Q)^2}{2n\Omega_n^2r_+^{2(n-1)}} + c_0c_1m^2r_+ + (n-1)c_0^2c_2m^2 \right. \nonumber\\
&&\left.+(n-1)k+\frac{(n-1)(n-2)c_0^3c_3m^2}{r_+} +\frac{(n-1)(n-2)(n-3)c_0^4c_4m^2}{r_+^2}\right].
\label{eq2.6}
\end{eqnarray}

\noindent The mass expressed by the horizon radius and charge is

\begin{eqnarray}
M &=& \frac{n\Omega_n r_+^{n-1}}{16\pi} \left[k+\frac{r_+^2}{l^2}+\frac{(16\pi Q)^2}{2n(n-1)\Omega_n^2r_+^{2(n-1)}}+\frac{c_0c_1m^2r_+}{n}+c_0^2c_2m^2 \right. \nonumber\\
&&\left.+\frac{(n-1)c_0^3c_3m^2}{r_+} +\frac{(n-1)(n-2)c_0^4c_4m^2}{r_+^2}\right].
\label{eq2.7}
\end{eqnarray}

\noindent The cosmological constant was seen as a fixed constant in the past. In this paper, it is regarded as a variable related to pressure, $P=-\frac{\Lambda}{8\pi} =\frac{n(n+1)}{16\pi l^2}$, and its conjugate quantity is a thermodynamic volume $V$. The entropy, volume and electric potential at the event horizon are given by

\begin{eqnarray}
S =\frac{\Omega_nr_+^{n}}{4}, \quad \quad\quad  V= \frac{\Omega_nr_+^{n+1}}{n+1},\quad \quad \quad \Phi_e= \frac{16\pi Q}{(n-1)\Omega_nr_+^{n-1}},
\label{eq2.8}
\end{eqnarray}

\noindent respectively. Due to the appearance of the pressure, the mass is no longer interpreted as the internal energy, but as an enthalpy. $c_1$, $c_2$, $c_3$ and $c_4$ are seen as extensive parameters for the mass. Their conjugate quantities are

\begin{eqnarray}
\Phi_1 &=& \frac{\Omega_nc_0m^2r_+^{n}}{16\pi},\quad \quad\quad \quad\quad\quad \Phi_2 = \frac{n\Omega_nc_0^2m^2r_+^{n-1}}{16\pi},\nonumber\\
\Phi_3 &=& \frac{n(n-1)\Omega_nc_0^3 m^2r_+^{n-2}}{16\pi}, \quad\quad  \Phi_4 = \frac{n(n-1)(n-2)\Omega_nc_0^4 m^2r_+^{n-3}}{16\pi},
\label{eq2.9}
\end{eqnarray}

\noindent respectively. It is easy to verify that these thermodynamic quantities obeys the first law of thermodynamics

\begin{eqnarray}
dM=TdS+VdP+\Phi_edQ+ \sum_{i=1}^4 \Phi_idc_i.
\end{eqnarray}

\noindent When the cosmological constant is fixed, the term $VdP$ disappears and the mass is interpreted as the internal energy. When a perturbative correction is introduced, the related thermodynamic quantities are shifted, which is discussed in the next section.

\section{Extremality relations in the massive gravity}

In this section, we derive the extremality relations between the mass and entropy, charge, pressure, parameters $c_i$ by adding a rescaled cosmological constant to the action as the perturbative correction. The rescaled parameter is $\epsilon$. Here, the black hole is designated as an extremal one.

We first introduce the correction

\begin{eqnarray}
\Delta \mathcal{S} &=& \frac{1}{16\pi}\int{dx^{n+2}\sqrt{-g}\frac{n(n+1)\epsilon}{l^2}},
\label{eq3.1}
\end{eqnarray}

\noindent to the action (\ref{eq2.1}). The corrected action is $\mathcal{S}+\Delta \mathcal{S}$. The action (\ref{eq2.1}) is recovered when $\epsilon = 0$. A black hole solution is gotten from the corrected action and takes the form as Eqs. (\ref{eq2.3}) and (\ref{eq2.5}), but there is a shift. Due to the correction, the Hawking temperature is also shifted and given by

\begin{eqnarray}
T &=& \frac{1}{4\pi r_+} \left[\frac{(n+1)(1+\epsilon)r_+^2}{l^2}+\frac{(16\pi Q)^2}{2n\Omega_n^2r_+^{2(n-1)}} + c_0c_1m^2r_+ + (n-1)c_0^2c_2m^2 \right. \nonumber\\
&&\left.+(n-1)k+\frac{(n-1)(n-2)c_0^3c_3m^2}{r_+} +\frac{(n-1)(n-2)(n-3)c_0^4c_4m^2}{r_+^2}\right].
\label{eq3.3}
\end{eqnarray}

\noindent The corrected mass is

\begin{eqnarray}
M &=& \frac{n\Omega_n r_+^{n-1}}{16\pi} \left[k+\frac{(1+\epsilon)r_+^2}{l^2}+\frac{(16\pi Q)^2}{2n(n-1)\Omega_n^2r_+^{2(n-1)}}+\frac{c_0c_1m^2r_+}{n}+c_0^2c_2m^2 \right. \nonumber\\
&&\left.+\frac{(n-1)c_0^3c_3m^2}{r_+} +\frac{(n-1)(n-2)c_0^4c_4m^2}{r_+^2}\right],
\label{eq3.2}
\end{eqnarray}

\noindent which is a function of parameters $r_+, \epsilon$, $Q, l, c_1, c_2, c_3$ and $c_4$. For convenience, we use $c$ to denote all parameters $Q, l, c_1, c_2, c_3, c_4$ except for $r_+$ and $\epsilon$. The differential of $M$ to $\epsilon$ is

\begin{eqnarray}
\left(\frac{\partial M}{\partial \epsilon}\right)_{c}&=& \left(\frac{\partial M}{\partial r_+}\right)_{c,\epsilon} \left(\frac{\partial r_+}{\partial \epsilon}\right)_{c} + \left(\frac{\partial M}{\partial \epsilon}\right)_{c,r_+}  \nonumber\\
&=& \left(\frac{\partial M}{\partial S}\right)_{c,\epsilon}\left(\frac{\partial S}{\partial r_+}\right)_{c,\epsilon} \left(\frac{\partial r_+}{\partial \epsilon}\right)_{c} + \left(\frac{\partial M}{\partial \epsilon}\right)_{c,r_+}\nonumber\\
&=& \frac{1}{4}T\Omega_nr_+^{n-1} \left(\frac{\partial r_+}{\partial \epsilon}\right)_{c} + \left(\frac{\partial M}{\partial \epsilon}\right)_{c, r_+}.
\end{eqnarray}

\noindent Our interest is focused on the thermodynamic extremality relation. In the extremal case, the temperature is zero, and the mass is rewritten as $M= M_{ext}$. Therefore, the above equation becomes

\begin{eqnarray}
\left(\frac{\partial M_{ext}}{\partial \epsilon}\right)_{Q, l, c_1, c_2, c_3, c_4}= \frac{n\Omega_nr_+^{n+1}}{16\pi l^2}.
\label{eq3.4}
\end{eqnarray}

\noindent Since the expression of the differential expressed by $\epsilon$ is very complex, we adopted the expression of $r_+$ in the above derivation. In fact, this relation can also be derived by a direct calculation. That the Hawking temperature (\ref{eq3.3}) is zero leads to a function $r_+ = r_+(\epsilon)$. Inserting this function into Eq. (\ref{eq3.2}) yields another function $M_{ext}= M_{ext}(\epsilon)$. Carrying out the differential on $M_{ext}(\epsilon)$ and using skills yield the same result as the relation (\ref{eq3.4}). The entropy $S$, pressure $P$, charge $Q$, $c_1$, $c_2$, $c_3$ and $c_4$ are usually regarded as a complete set of extensive parameters for the mass. Their conjugate quantities can be derived from the mass and take the same form as those given in section 2 except for the temperature and volume. We first verify the extremality relation between the mass and entropy.

From Eq. (\ref{eq3.2}), the expression of $\epsilon$ is gotten and takes the form

\begin{eqnarray}
\epsilon &=&\left[\frac{16\pi M}{n\Omega_n r_+^{n+1}} -\frac{k}{r_+^2} -\frac{(16\pi Q)^2}{2n(n-1)\Omega_n^2r_+^{2n}} -\frac{c_0c_1m^2}{nr_+} - \frac{c_0^2c_2m^2}{r_+^2} \right. \nonumber\\
&&\left.-\frac{(n-1)c_0^3c_3m^2}{r_+^3} -\frac{(n-1)(n-2)c_0^4c_4m^2}{r_+^4}\right] l^2-1.
\label{eq3.5}
\end{eqnarray}

\noindent Due to the relation between the entropy and horizon radius given in Eq. $(\ref{eq2.8})$, the above equation is a function $\epsilon (S)$ and $\frac{\partial r_+}{\partial S}=\frac{4}{n\Omega_n r_+^{n-1}}$. Carrying out the differential calculation on this function yields

\begin{eqnarray}
\left(\frac{\partial \epsilon}{\partial S}\right)_{M,Q, l, c_1, c_2, c_3, c_4}&=& \frac{4l^2}{n\Omega_n r_+^{n-1}}\left[-\frac{(n+1)16\pi M}{n\Omega_n r_+^{n+2}} +\frac{2k}{r_+^3} +\frac{(16\pi Q)^2}{(n-1)\Omega_n^2r_+^{2n+1}} \right. \nonumber\\
&& +\frac{c_0c_1m^2}{nr_+^2}+ \frac{2c_0^2c_2m^2}{r_+^3} +\frac{3(n-1)c_0^3c_3m^2}{r_+^4}\nonumber\\
&&\left. +\frac{4(n-1)(n-2)c_0^4c_4m^2}{r_+^5}\right].
\label{eq3.6}
\end{eqnarray}

\noindent To evaluate the value, we insert the expression of the mass into the above equation and get

\begin{eqnarray}
\left(\frac{\partial \epsilon}{\partial S}\right)_{M,Q, l, c_1, c_2, c_3, c_4}&=& \frac{4l^2}{n\Omega_n r_+^{n-1}}\left[-\frac{(n-1)k}{r_+^3} -\frac{(n+1)(1+\epsilon)}{r_+l^2} + \frac{(16\pi Q)^2}{2n\Omega_n^2r_+^{2(n+1)}} \right. \nonumber\\
&& - \frac{c_0c_1m^2}{r_+^2} - \frac{(n-1)c_0^2c_2m^2}{r_+^3} -\frac{(n-1)(n-2)c_0^3c_3m^2}{r_+^4}\nonumber\\
&&\left. -\frac{(n-1)(n-2)(n-3)c_0^4c_4m^2}{r_+^5}\right].
\label{eq3.7}
\end{eqnarray}

\noindent Combining the inverse of the above differential with the expression of the temperature given in Eq. (\ref{eq3.3}), we have

\begin{eqnarray}
T\left(\frac{\partial S}{\partial \epsilon}\right)_{M,Q, l, c_1, c_2, c_3, c_4}=-\frac{n\Omega_nr_+^{n+1}}{16\pi l^2}.
\label{eq3.8}
\end{eqnarray}

\noindent Compared it with the relation (\ref{eq3.4}), it is easy to get

\begin{eqnarray}
\left(\frac{\partial M_{ext}}{\partial \epsilon}\right)_{Q, l, c_1, c_2, c_3, c_4}= \lim_{M \to M_{ext}} - T\left(\frac{\partial S}{\partial \epsilon}\right)_{M,Q, l, c_1, c_2, c_3, c_4},
\label{eq3.9}
\end{eqnarray}

\noindent where $S$ is a function of $M$, $Q$, $l$, $c_1$, $c_2$, $c_3$, $c_4$ and $\epsilon$. Therefore, the Goon-Penco relation is verified in the higher-dimensional black hole.

In this paper, the cosmological constant is regarded as a variable related to pressure. The entropy, pressure, charge, $c_1$, $c_2$, $c_3$ and $c_4$ are usually regarded as the extensive parameters for the mass. Since the entropy satisfies the thermodynamic extremality relation, it is natural to ask whether other extensive quantities also satisfy corresponding relations. What we need to do in the following investigation is to find out these relations. Let's first derive the extremality relation between the mass and pressure. The pressure is expressed by the constant $l^2$ as $P = \frac{n(n+1)}{16\pi l^2}$. Then, $\frac{\partial P}{\partial l^2}=-\frac{16\pi l^4}{n(n+1)}$. Using Eqs. (\ref{eq3.2}) and (\ref{eq3.5}), we get the differential of $\epsilon$ to the pressure,

\begin{eqnarray}
\left(\frac{\partial \epsilon}{\partial P}\right)_{M, r_+, Q, c_1, c_2, c_3, c_4} = \frac{-16\pi l^2(1+\epsilon)}{n(n+1)}.
\label{eq3.15}
\end{eqnarray}

\noindent The perturbation parameter $\epsilon$ exists in the above differential relation as an explicit function. The reason is that the perturbation correction is introduced by adding the rescaled cosmological constant to the action and this constant is related to the pressure. Due to the shift of the mass, the thermodynamic volume is also shifted and its expression is different from that given given in Eq. (\ref{eq2.8}). The volume is

\begin{eqnarray}
V=\frac{\epsilon +1}{n+1}\Omega_nr_+^{n+1}.
\label{eq3.16}
\end{eqnarray}

\noindent Using Eq. (\ref{eq3.16}) and the inverse of the differential of $\epsilon$ to $P$ yields

\begin{eqnarray}
V\left(\frac{\partial P}{\partial \epsilon}\right)_{M, r_+, Q, c_1, c_2, c_3, c_4}=-\frac{n\Omega_nr_+^{n+1}}{16\pi l^2}.
\label{eq3.17}
\end{eqnarray}

\noindent Comparing the above equation with Eq. (\ref{eq3.4}), we get the extremality relation between the mass and pressure,

\begin{eqnarray}
\left(\frac{\partial M_{ext}}{\partial \epsilon}\right)_{Q, l, c_1, c_2, c_3, c_4}= \lim_{M \to M_{ext}} - V\left(\frac{\partial P}{\partial \epsilon}\right)_{M, r_+, Q, c_1, c_2, c_3, c_4},
\label{eq3.18}
\end{eqnarray}

\noindent where $P$ is a function of $M$, $r_+$, $Q$, $c_1$, $c_2$, $c_3$, $c_4$ and $\epsilon$. This relation is an extension of the Goon-Penco relation.

We continue to investigate the extremality relation between the mass and charge. The calculation process is similar. From Eq. (\ref{eq3.5}), the differential of $\epsilon$ to $Q$ takes the form

\begin{eqnarray}
\left(\frac{\partial \epsilon}{\partial Q}\right)_{M, r_+, l, c_1, c_2, c_3, c_4} = - \frac{(16\pi)^2 Ql^2}{n(n-1)\Omega_n^2r_+^{2n}}.
\label{eq3.10}
\end{eqnarray}

\noindent Multiplying the electric potential $\Phi_e=\frac{16\pi Q}{(n-1)\Omega_nr_+^{n-1}}$ by the inverse of the above differential yields

\begin{eqnarray}
\Phi_e\left(\frac{\partial Q}{\partial \epsilon}\right)_{M, r_+, l, c_1, c_2, c_3, c_4}=-\frac{n\Omega_nr_+^{n+1}}{16\pi l^2}.
\label{eq3.11}
\end{eqnarray}

\noindent Obviously, there is a minus sign difference between Eqs. (\ref{eq3.4}) and (\ref{eq3.11}). Therefore,

\begin{eqnarray}
\left(\frac{\partial M_{ext}}{\partial \epsilon}\right)_{Q, l, c_1, c_2, c_3, c_4}= \lim_{M \to M_{ext}} - \Phi_e\left(\frac{\partial Q}{\partial \epsilon}\right)_{M, r_+, l, c_1, c_2, c_3, c_4},
\label{eq3.12}
\end{eqnarray}

\noindent which is the extremality relation between the mass and charge. Now, $Q$ is a function of $M$, $r_+$, $l$, $c_1$, $c_2$, $c_3$, $c_4$ and $\epsilon$. This relation is also an extension of the Goon-Penco relation.

For the extremality relations between the mass and parameters $c_1$, $c_2$, $c_3$, $c_4$, the calculations are parallel. Their differential relations are

\begin{eqnarray}
\left(\frac{\partial \epsilon}{\partial c_1}\right)_{M, r_+, Q, l, c_2, c_3, c_4} &=& - \frac{c_0m^2l^2}{nr_+},\\
\left(\frac{\partial \epsilon}{\partial c_2}\right)_{M, r_+, Q, l, c_1, c_3, c_4} &=& - \frac{c_0^2m^2l^2}{r_+^2},\\
\left(\frac{\partial \epsilon}{\partial c_3}\right)_{M, r_+, Q, l, c_1, c_2, c_4} &=& - \frac{(n-1)c_0^3m^2l^2}{r_+^3},\\
\left(\frac{\partial \epsilon}{\partial c_4}\right)_{M, r_+, Q, l, c_1, c_2, c_3} &=& - \frac{(n-1)(n-2)c_0m^2l^2}{r_+^4}.
\end{eqnarray}

\noindent The conjugate quantities of $c_1$, $c_2$, $c_3$ and $c_4$ are $\Phi_1 = \frac{\Omega_nc_0m^2r_+^{n}}{16\pi}$, $\Phi_2 = \frac{n\Omega_nc_0^2m^2r_+^{n-1}}{16\pi}$, $\Phi_3 = \frac{n(n-1)\Omega_nc_0^3 m^2r_+^{n-2}}{16\pi}$ and $\Phi_4 = \frac{n(n-1)(n-2)\Omega_nc_0^4 m^2r_+^{n-3}}{16\pi}$, respectively. Using these quantities, it is not difficult for us to get

\begin{eqnarray}
&&\left(\Phi_1\frac{\partial c_1}{\partial \epsilon}\right)_{M, r_+, Q, l, c_2, c_3, c_4} =\left(\Phi_2\frac{\partial c_2}{\partial \epsilon}\right)_{M, r_+, Q, l, c_1, c_3, c_4}=\left(\Phi_3\frac{\partial c_3}{\partial \epsilon}\right)_{M, r_+, Q, l, c_1, c_2, c_4} \nonumber\\
&& =\left(\Phi_4\frac{\partial c_4}{\partial \epsilon}\right)_{M, r_+, Q, l, c_1, c_2, c_3}=-\frac{n\Omega_nr_+^{n+1}}{16\pi l^2}.
\label{eq3.13}
\end{eqnarray}

\noindent Thus, the extremality relations between the mass and extensive parameters $c_i$ are

\begin{eqnarray}
\left(\frac{\partial M_{ext}}{\partial \epsilon}\right)_{Q, l, c_1, c_2, c_3, c_4}= \lim_{M \to M_{ext}} - \Phi_i\left(\frac{\partial c_i}{\partial \epsilon}\right)_{M, r_+, Q, l, c_j,c_k, c_u},
\label{eq3.14}
\end{eqnarray}

\noindent where $i,j,k,u=1,2,3,4$ and $i \neq j \neq k \neq u$. Therefore, the Goon-Penco relation is extended to the case of the extensive parameters $c_i$ of the higher-dimensional black hole.

In the above investigation, the thermodynamic extremality relations between the mass and entropy, pressure, charge, parameters $c_i$ were gotten by the accurate calculations. They are expressed by Eqs. (\ref{eq3.8}), (\ref{eq3.18}), (\ref{eq3.12}) and (\ref{eq3.14}), respectively. The values of these relations are equal. The result shows that Goon-Penco relation is extended to the higher-dimensional black hole in the massive gravity.

\section{Discussion and conclusion}

In this paper, we extended the work of Goon and Penco to the massive gravity and investigated the thermodynamic extremality relations in the higher-dimensional black hole. The perturbative correction was introduced by adding the rescaled cosmological constant to the action. The extremality relations between the mass and pressure, entropy, charge, parameters $c_i$ were derived by the accurate calculations, respectively. The values of these extremality relations are equal, which may be due to the first law of thermodynamics.
In the calculation, the cosmological constant was seen as a variable related to pressure. Its conjugate quantity is a thermodynamic volume. Although the rescaled constant was added to the action, this addition does not affect the form of the extremality relation between the mass and pressure.

In fact, these relations can be derived uniformly by using the triple product identity

\begin{eqnarray}
\left(\frac{\partial M}{\partial X^i}\right)_{\epsilon, T}\left(\frac{\partial X^i}{\partial \epsilon}\right)_{M,T} \left(\frac{\partial \epsilon}{\partial M}\right)_{T, X^i}=-1,
\label{eq4.1}
\end{eqnarray}

\noindent which yields

\begin{eqnarray}
\left(\frac{\partial M}{\partial \epsilon}\right)_{T, X^i}=-\left(\frac{\partial M}{\partial X^i}\right)_{\epsilon, T}\left(\frac{\partial X^i}{\partial \epsilon}\right)_{M,T}=-\Phi_i\left(\frac{\partial X^i}{\partial \epsilon}\right)_{M,T}.
\label{eq4.2}
\end{eqnarray}

\noindent In the above derivation, $\left(\frac{\partial M}{\partial X^i}\right)_{\epsilon, T}$ were identified to $\Phi_i$ which are the conjugate quantities to $X^i$. For the charged black hole given in section 3, $X^i$ are chosen as $S$, $Q$, $P$, $c_1$, $c_2$, $c_3$ and $c_4$. $M$ and $T$ are the corrected mass and temperature given in (\ref{eq3.2}) and (\ref{eq3.3}), respectively. In the extremal case, $T \to 0$ and $M \to M_{ext}$. The above relation becomes

\begin{eqnarray}
\left(\frac{\partial M_{ext}}{\partial \epsilon}\right)_{M,X^{i}}=  \lim_{M \to M_{ext}}-\Phi_i\left(\frac{\partial X^i}{\partial \epsilon}\right)_{M,X^{j}},
\label{eq4.3}
\end{eqnarray}

\noindent where $X^j$ are chosen as $S$, $Q$, $P$, $c_1$, $c_2$, $c_3$ and $c_4$, but $X^j \neq X^j$. This relation implies that the universal extremality relation exist in black holes. The relation (\ref{eq4.3}) is easily reduced to (\ref{eq3.8}), (\ref{eq3.18}), (\ref{eq3.12}) and (\ref{eq3.14}) when $X^i$ is the entropy, charge, parameters $c_i$ and pressure. In the calculation, due to the shift of the mass, the expression of the volume $V=\frac{\epsilon +1}{n+1}\Omega_nr_+^{n+1}$ is different from that given in Eq. (\ref{eq2.8}). In \cite{WYL}, the authors derived the extremality relation between the mass and angular momentum in the BTZ and Kerr anti-de Sitter spacetimes, and made a conjecture that a general formula of the extremality relation existed in black holes. Our result gives a verification to this conjecture.

\begin{acknowledgments}
This work is supported by NSFC (Grant Nos. 11875196, 11375121, 11205125 and 11005016) and FXHU (Z201021).
\end{acknowledgments}


\begin{thebibliography}{99}                                                                                               %

\bibitem{CV}
C. Vafa, \emph{The string landscape and the swampland,} [arXiv:hep-th/0509212].

\bibitem{AMNV}
N. Arkani-Hamed, L. Motl, A. Nicolis and C. Vafa, \emph{The string landscape, black holes and gravity as the weakest force}, \emph{JHEP} \textbf{0706} (2007) 060.

\bibitem{EP}
E. Palti, \emph{The swampland: introduction and review}, \emph{Fortsch. Phys.} \textbf{67} (2019) 1900037.

\bibitem{BCV}
T.D. Brennan, F. Carta and C. Vafa, \emph{The string landscape, the swampland, and the missing cornerw}, \emph{PoS TASI} \textbf{2017} (2017) 015.

\bibitem{CLR1}
C. Cheung, J. Liu, and G.N. Remmen, \emph{Proof of the weak gravity conjecture from black hole entropy}, \emph{JHEP} \textbf{10} (2018) 004.

\bibitem{CLR2}
C. Cheung, J. Liu, and G. N. Remmen, \emph{Entropy bounds on effective field theory from rotating dyonic black holes}, \emph{Phys. Rev.} \textbf{D 100} (2019) 046003.

\bibitem{JM}
C.R.T. Jones and B. McPeak, \emph{The black hole weak gravity conjecture with multiple charges}, [arXiv:1908.10452[th-hep]]

\bibitem{CHNW}
W.M. Chen, Y.T. Huang, T. Noumi and C.K. Wen, \emph{Unitarity bounds on charged/neutral state mass ratio}, \emph{Phys. Rev.} \textbf{D 100} (2019) 025016.

\bibitem{ACS}
L. Aalsma, A. Cole and G. Shiu, \emph{Weak gravity conjecture, black hole entropy, and modular invariance}, \emph{JHEP} \textbf{1908} (2019) 022.

\bibitem{LLW1}
S.J. Lee, W. Lerche and T. Weigand, \emph{A stringy test of the scalar weak gravity conjecture}, \emph{Nucl. Phys.} \textbf{B 938} (2019) 321

\bibitem{LLW2}
S.J. Lee, W. Lerche and T. Weigand, \emph{Modular fluxes, elliptic genera, and weak gravity conjectures in four dimensions}, \emph{JHEP} \textbf{1908} (2019) 104.

\bibitem{KMP}
Y. Kats, L. Motl and M. Padi, \emph{Higher-order corrections to mass-charge relation of extremal black holes}, \emph{JHEP} \textbf{0712} (2007) 068.

\bibitem{KMP1}
K. Kooner, S. Parameswaran and I. Zavala, \emph{Warping the weak gravity conjecture}, \emph{Phys. Lett.} \textbf{B 759} (2016) 402.

\bibitem{KMP2}
E. Gonzalo and L.E. Ib$\acute{a}$$\tilde{n}$ez, \emph{The fundamental need for a SM Higgs and the weak gravity conjecture}, \emph{Phys. Lett.} \textbf{B 786} (2018) 272.

\bibitem{LNS}
G.J. Loges, T. Noumi and G. Shiu, \emph{Thermodynamics of 4D Dilatonic black holes and the weak gravity conjecture}, [arXiv:1909.01352 [hep-th]].

\bibitem{MRV}
M. Montero, T.V. Riet and G. Venken, \emph{Festina lente: EFT constraints from charged black hole evaporation in de Sitter}, \emph{JHEP} \textbf{2001} (2020) 039.

\bibitem{CCLOR}
P.A. Cano, S. Chimento, R. Linares, T. Ortin and P.F. Ramirez, \emph{$\alpha ^{\prime}$ corrections of Reissner-Nordstr$\ddot{o}$m black holes}, \emph{JHEP} \textbf{2002} (2020) 031.

\bibitem{COR}
P.A. Cano, T. Ortin and P.F. Ramirez, \emph{On the extremality bound of stringy black holes}, \emph{JHEP} \textbf{2002} (2020) 175.

\bibitem{RS}
H.S. Reall and J.E. Santos, \emph{Higher derivative corrections to Kerr black hole thermodynamics}, \emph{JHEP} \textbf{1904} (2019) 021.

\bibitem{HRR}
B. Heidenreich, M. Reece and T. Rudelius, \emph{Repulsive forces and the weak gravity conjecture}, \emph{JHEP} \textbf{1910} (2019) 055.

\bibitem{BH}
S. Brahma and M.W. Hossain, \emph{Relating the scalar weak gravity conjecture and the swampland distance conjecture for an accelerating universe}, \emph{Phys. Rev.} \textbf{D 100} (2019) 086017.

\bibitem{GI}
E. Gonzalo, L.E. Ib$\acute{a}\tilde{n}$ez, \emph{A strong scalar weak gravity conjecture and some implications}, \emph{JHEP} \textbf{1908} (2019) 118.

\bibitem{HNS}
Y. Hamada, T. Noumi and G. Shiu, \emph{Weak gravity conjecture from unitarity and causality}, \emph{Phys. Rev. Lett.} \textbf{123} (2019) 051601.

\bibitem{GP}
G. Goon and R. Penco, \emph{A universal relation between corrections to entropy and extremalitys}, \emph{Phys. Rev. Lett.} \textbf{124} (2020) 101103.

\bibitem{CJLM}
S. Cremonini, C.R.T. Jones, J.T. Liu and B. McPeak, \emph{Higher-derivative corrections to entropy and the weak gravity conjecture in anti-de Sitter space}, [arXiv:1912.11161 [hep-th]].

\bibitem{WYL}
S.W. Wei, K. Yang and Y.X. Liu, \emph{Universal thermodynamic relations with constant corrections for rotating AdS black holes}, [arXiv:2003.06785 [gr-qc]].

\bibitem{KRT}
D. Kastor, S. Ray and J. Traschen, \emph{Enthalpy and the mechanics of AdS black holes}, \emph{Class. Quant. Grav.} \textbf{26} (2009) 195011.

\bibitem{KRT1}
D. Kastor, S. Ray and J. Traschen, \emph{Smarr formula and an extended first law for lovelock gravity}, \emph{Class. Quantum Gravity} \textbf{27} (2010) 235014.

\bibitem{LPVP}
H. Lu, Y. Pang, C.N. Pope and J.F. Vazquez-Poritz, \emph{AdS and Lifshitz black holes in conformal and Einstein-Weyl gravities}, \emph{Phys. Rev.} \textbf{D 86} (2012) 044011.

\bibitem{KM1}
D. Kubiznak and R.B. Mann, \emph{P-V criticality of charged AdS black holes}, \emph{JHEP} \textbf{1207} (2012) 033.

\bibitem{CM}
J. Creighton and R.B. Mann, \emph{Quasilocal thermodynamics of dilaton gravity coupled to gauge fields}, \emph{Phys. Rev.} \textbf{D 52} (1995) 4569.

\bibitem{DV}
D. Vegh, \emph{Holography without translational symmetry}, [arXiv:1301.0537 [hep-th]].

\bibitem{CHPZ}
R.G. Cai, Y.P. Hu, Q.Y. Pan and Y.L. Zhang, \emph{Thermodynamics of black holes in massive gravity}, \emph{Phys. Rev.} \textbf{D 91} (2015) 024032.

\bibitem{DYC}
D.Y. Chen, \emph{Thermodynamics and weak cosmic censorship conjecture in extended phase spaces of anti-de Sitter black holes with particles' absorption}, \emph{Eur. Phys. J.} \textbf{C 79} (2019) 353.

\bibitem{XCH1}
J.F. Xu, L.M. Cao and Y.P. Hu, \emph{$P-V$ criticality in the extended phase space of black holes in massive gravity}, \emph{Phys. Rev.} \textbf{D 91} (2015) 124033.

\bibitem{XCH2}
D.C. Zou, R.H. Yue and M. Zhang, \emph{Behavior of quasinormal modes and Van der Waals-like phase transition of charged AdS black holes in massive gravity}, \emph{Eur. Phys. J.} \textbf{C 77} (2017) 256.

\bibitem{XCH3}
D.C. Zou, Y.Q. Liu and R.H. Yue, \emph{Behavior of quasinormal modes and Van der Waals-like phase transition of charged AdS black holes in massive gravity}, \emph{Eur. Phys. J.} \textbf{C 77} (2017) 365.







\end{thebibliography}
\end{document}